\documentstyle[aps,epsfig]{revtex}
\newcommand{\be}{\begin{equation}}
\newcommand{\ee}{\end{equation}}
\newcommand{\bea}{\begin{eqnarray}}
\newcommand{\eea}{\end{eqnarray}}
\newcommand{\hf} {{1\over2}}
\newcommand{\nonu}{\nonumber\\}
\def\eq#1{(\ref{#1})}
\def\la{\langle}
\def\ra{\rangle}
\def\cd#1{{\cal D}[#1]}
\def\cD{{\cal D}}
\def\ide{{1\over\Delta}}

\def\tcG{\tilde{\cal G}}
\def\tcGi{\tilde{\cal G}^{-1}}
\def\psid{\psi^\dagger}
\def\tr{{\mathrm Tr}}
\def\ve{V^{\mathrm ext}}
\def\ord#1{{\cal O}(#1)}
\def\tGi{{\tilde G}^{-1}}
\def\rt{\rho^*}
\def\cC{{\cal C}} 
\def\mb#1{{\mathbf#1}}
\def\mr#1{{\mathrm#1}}
\def\tG{\tilde G}

\begin{document}
\title{INTERNAL SPACE RENORMALIZATION GROUP METHODS FOR ATOMIC AND
CONDENSED MATTER PHYSICS}
\author{Janos Polonyi}
\address{Laboratory of Theoretical Physics,
Louis Pasteur University, Strasbourg, France,\\
Department of Atomic Physics, Lorand E\"otv\"os University,
Budapest, Hungary}
\maketitle
\begin{abstract}
The functional renormalization group method is used to
take into account the vacuum polarization around localized bound 
states generated by external potential. The application to 
Atomic Physics leads to improved Hartree-Fock and Kohn-Sham 
equations in a systematic manner within the framework of the 
Density Functional Theory. Another application to Condensed Matter 
Physics consists of an algorithm to compute quenched averages with 
or without Coulomb interaction in a non-perturbative manner.
\end{abstract}
\pacs{31.15.-p, 72.10.-d}
\date{\today}
\section{Introduction}
The renormalization group will be used in this talk\footnote{deliviered at
the Conference {\bf Renormalization
Group 2002 (RG-2002)} Strba, Slovakia, March 2002} as an algorithm to
solve strongly coupled quantum field theories without any intention
to gain insight into the scale dependence of the dynamics. We shall
study the system of non-relativistic electrons propagating in the presence of
static external potential. We consider non-relativistic
systems because the comparison with experiment is more direct than in
the relativistic domain. The common challenge in both regimes is to
trace the polarization effects in the vacuum, i.e. the particle-hole
or the particle-anti particle fluctuations in the non-relativistic or 
relativistic region, respectively.

We shall consider two different cases. First, the electrons will 
be placed in an external localized field and one finds a problem 
with inhomogeneous ground state, characteristic of Atomic Physics .
Second, we assume the presence of a static random external impurity
potential which obeys a Gaussian probability distribution and we introduce
a scheme to compute the quenched averages of kinetic transport coefficients
in a translation invariant manner. The common technics applied is a 
generalization of the functional renormalization group \cite{func}
where the role of the running cut-off is played by an arbitrary
control parameter which generates differentiable changes in the dynamics.

\section{Renormalization in the internal space}
The degrees of freedom are eliminated successively in the
renormalization group method, for instance the Kadanoff-Wilson blocking
strategy orders the modes according to their scales in the external space, 
i.e. in the the space-time (energy-momentum). One may construct a blocking
procedure where the blocking proceeds
in an order determined by a scale in the internal space, the
space of the field amplitude. A well known example is the
Callan-Symanzik equation where the fluctuations with larger and 
larger amplitudes are taken into account as the mass of the 
particles is lowered.

It is easy to generalize
this scheme for any parameter in the dynamics by means of 
functional techniques \cite{int}. In fact, let us suppose that the
dynamics contains a control parameter $\lambda$ and the
generator functional for the connected Green functions is given by
\be
e^{{i\over\hbar}W_\lambda[j]}=\int\cD[\phi]e^{{i\over\hbar}
(S_\lambda[\phi]+j\cdot\phi)}
\ee
where $f\cdot g=\int dxf_xg_x$. The evolution equation
\be\label{evole}
e^{{i\over\hbar}W_\lambda[j]}\partial_\lambda W_\lambda[j]=
\int\cD[\phi]\partial_\lambda S_\lambda[\phi]e^{{i\over\hbar}
(S_\lambda[\phi]+j\cdot\phi)}
=\partial_\lambda S_\lambda\left[
{\hbar\over i}{\delta\over\delta j}\right]e^{{i\over\hbar}W_\lambda[j]}
\ee
is valid without assuming the existence of any small parameter. 
The way this equation is obtanied is reminescent of the 
derivation of the Schwinger-Dyson equations except that
the latter contains the complete action and the former 
the suppression controling part, $\partial_\lambda S_\lambda$, only. It is 
worthwhile noting that this relation appears to be such a functional 
generalization of the Hellman-Feynman theorem \cite{helfey},
\be
\partial_\lambda E_\lambda
=\la E_\lambda|\partial_\lambda H_\lambda|E_\lambda\ra,~~~
E_\lambda|E_\lambda\ra=H_\lambda|E_\lambda\ra,
\ee
for time dependent processes and general matrix elements which
remains compatible with approximations one employs to solve the
functional differential equation \eq{evole} in restricted functional
spaces.

In the examples presented here the
electric charge and the average strength of the impurity
field fluctuations will be chosen as control parameters.
The corresponding evolution equations will be written for a local
functional, the effective action
for the density and the current. The effective action of the
physical system can be obtained by integrating these equations
from an artificial perturbative initial condition imposed at weak 
Coulomb interaction and disorder into the physical regime.

Similar schemes could be constructed by means of the more traditional
renormalization group method, based on the external space. 
But the present version has the 
following two advantages. First, it preserves gauge invariance, an important 
feature in the computation of the electric conductivity. Second, it
avoids the artificial discontinuities during the evolution 
which may occur when the saddle point structure is changed by the
running cut-off \cite{tree}.

\section{Density Functional Theory}
The density functional theory \cite{hoko,ksh} is a powerful 
method to describe bound states induced by external potential. 
The quantity of central importance
is the density functional, $E_\mr{gr}[\rho]$, the ground state energy when 
the the density is constrained to be $\rho$. According to the
the Hohenberg-Kohn theorems the ground state energy of the original
problem is the minimum of the density functional,
$E_\mr{gr}=E_\mr{gr}[\rho_\mr{gr}]$. The shortcoming of the traditional
approach is its phenomenological nature, the lack of a constructive 
definition of the density functional. We propose that the effective
action $\Gamma[\rho]$ for the density $\rho$ \cite{efa} 
provides a clear and generally applicable definition of the density functional.
Furthermore, the internal space renormalization group method leads to a 
systematical approximation scheme which turns the phenomenological 
knowledge about correlations into an improvement of the approximation
by choosing the ansatz for $\Gamma[\rho]$ in an appropriate manner
\cite{kornel}.

The generator functional for the connected Green function is defined as
\be\label{congrgf}
e^{{i\over\hbar}W[\sigma]}=\int\cd{u}\cd{\psi}\cd{\psid}
e^{{i\over\hbar}\psid\cdot(G^{-1}+\sigma_\alpha\cdot j_\alpha)\cdot\psi
+{i\over2\hbar}\partial u\cdot\partial u},
\ee
where $j_{\alpha,x}$ is the electric density ($\alpha=0$) and current
($\alpha=1,2,3$). We shall consider the density Green functions 
and use $\alpha=0$ only in this section. The Legendre transform of the
$W[\sigma]$, the effective action $\Gamma[\rho]$ for the density $\rho$
reaches its minimum at the
ground state density, $\beta E_\mr{gr}=\Gamma[\rho_\mr{gr}]$, and
will serve as the density functional.
The effective action is first obtained for imaginary time,
at finite temperature, $T=1/\beta$,
and the projection onto the ground state is achieved by taking the
zero temperature limit $\beta\to\infty$.
After the rescaling $e\to\lambda e$ of the electric charge one can derive the
evolution equation
\bea\label{cevol}
\partial_\lambda\Gamma_\lambda[\rho]&=&
-\lambda e^2\left\{\rho\cdot\ide\cdot\rho+\tr\left[
\left({\delta^2\Gamma_\lambda[\rho]\over\delta\rho\delta\rho}\right)^{-1}
\ide\right]\right\}\nonu
&=&\int_{\mb{x},\mb{y},t}\la\psid_{\mb{x},t}\psi_{\mb{x},t}
\psid_{\mb{y},t}\psi_{\mb{y},t}\ra\partial_\lambda{\lambda^2e^2
\over8\pi|\mb{x}-\mb{y}|},
\eea
where $\psi$ denotes the electron field operator and $\int_x=\int dx$.
The second equation shows that the evolution equation simply follows
the change of the Coulomb energy according to the Hellman-Feynman theorem.

The initial condition for the effective action is imposed at weak
Coulomb interaction, $\lambda_0\approx0$, where we find in the leading
order perturbation expansion for spinless electrons 
\be
\Gamma[\rho]=\hf(\rho-\rt)\cdot\tcGi\cdot(\rho-\rt)
-\rho\cdot{1\over2\Delta}\cdot\rho+\cC[\rho]
\ee
with $\cC[\rho]=-\tr\log G^{-1}+\hf\tr\log D^{-1}+\ord{e^2}+\ord{\rho^4}$.
Here 
\be
G^{-1}_{x,x'}=\delta_{x,x'}\left(\partial_t
-{\hbar^2\over2m}\Delta_\mb{x}-\mu+\ve_\mb{x}\right)
\ee
is the propagator for non-interacting electrons in the presence of an 
external potential $\ve$ and chemical potential $\mu$, 
$\tG_{x,y}=-G_{x,y}G_{y,x}$ stands for the particle-hole propagator, 
$\tcG=[\tGi+e^2/\Delta]^{-1}$ is its improved one-lop version, 
$D=[-\Delta+e^2\tG]^{-1}$ denotes
the photon propagator and $\rt_x=-G_{x,x}$. By minimizing the effective
action one recovers the Hartree-Fock energy functional in $\ord{e^2}$.
The non-classical nature of the exchange contribution to the interaction 
is reflected in the fact that it arises from the photon fluctuation 
determinant, the term $\tr\log D^{-1}$.

The evolution equation \eq{cevol} should be projected into a restricted
functional space in order to make it more manageable. For this end
we introduce the multi-local truncation scheme for the effective action.
The free $k$-local cluster approximation, $f_k$, corresponds to the 
functional form
\be
\Gamma[\rho]=\sum_{j=0}^k\prod_{j=1}^k\sum_{n_j}\int dx_j
\Gamma^{n_1,\ldots,n_k}_{x_1,\ldots,x_k}\rho_{x_1}^{n_1}\cdots\rho_{x_k}^{n_k}
\ee
with arbitrary $\Gamma^{n_1,\ldots,n_k}_{x_1,\ldots,x_k}$.
The effective action in the constrained $k$-local cluster approximation, 
$c_k$, can be written in the same manner except that the functions
$\Gamma^{n_1,\ldots,n_k}_{x_1,\ldots,x_k}$ are parameterized. The
interactive electron field operator is assumed to be of the form
\be\label{sqop}
\psi_{t,\mb{x}}=\sum_{n=1}^\infty c_{n,t}\Psi_{n,\mb{x}},
\ee
and the propagator will be written as
\be\label{fgrspi}
G_{x,x'}=
\sum_{n=1}^Ne^{-E_n(t-t')}\Psi_{n,\mb{x}}\Psi^*_{n,\mb{x'}}
+\sum_{n,n'=1}^\infty g_{n,n'}(t-t')\Psi_{n,\mb{x}}\Psi^*_{n',\mb{x'}},
\ee
for $t<t'$ leaving $\Psi_{n,\mb{x}}$, $E_n$ and $g_{n,n'}(t)$ as
parameters.

We introduce a local density-dependent self-energy, $\sigma(x,\rho_x)$,
in the photon propagator and assume the form
\be
\cC[\rho]=-\tr\log G^{-1}+\hf\tr\log[D^{-1}+\sigma]
+\int_xU(x,\rho_x)+\sum_{n,m}\rho^n\cdot\gamma^{(n,m)}\cdot\rho^m
\ee
which is a $f_2$ ansatz with infinitely many higher order constrained 
clusters arising from the photon fluctuation determinant. The corresponding
evolution equation, considered at the minimum of the effective action,
\be
\rho_\mr{gr}=\left(1+\tG\cdot{\lambda^2e^2\over\Delta}\right)\cdot
\left(\rt-\tcG\cdot{\delta\cC[\rho_\mr{gr}]\over\delta\rho}\right),
\ee
 can be written in a variational form as
\be\label{vareveq}
{\delta{\cal H}[\Psi^*,\Psi,E,\rho]\over\delta\Psi^*_{n,\mb{x}}}
_{\vert\rho=\rho_\mr{gr}}
={\delta{\cal H}[\Psi^*,\Psi,E,\rho]\over\delta\Psi_{n,\mb{x}}}
_{\vert\rho=\rho_\mr{gr}}
={\partial{\cal H}[\Psi^*,\Psi,E,\rho]\over
\partial E_n}_{\vert\rho=\rho_\mr{gr}}=0,
\ee
after the appropriate choice of a constant $\Gamma_0$. A generalized HF 
functional
\be\label{hffunc}
{\cal H}[\Psi^*,\Psi,E,\rho]={\cal H}^\mr{fr}[\Psi^*,\Psi,E]
+{\cal H}^\mr{C}[\Psi^*,\Psi]
+{\cal H}^\mr{ph}[\Psi^*,\Psi,\rho]
+{\cal H}^\mr{i}[\Psi^*,\Psi,\rho],
\ee
was introduced here which is the sum of the one-particle, exchange, direct 
and interaction pieces,
\bea
{\cal H}^\mr{fr}[\Psi^*,\Psi,E]&=&\sum_{n=1}^N\left[E_n+\int_\mb{x}
\Psi_{n,\mb{x}}^*\left(-E_n-{\hbar^2\over2m}\Delta
+\ve_\mb{x}\right)\Psi_{n,\mb{x}}\right],\nonu
{\cal H}^\mr{ph}[\Psi^*,\Psi,\rho]&=&{1\over2\beta}\tr\log
\left(-\Delta+\sigma(\rho)+\lambda^2e^2\tG\right),\nonu
{\cal H}^\mr{C}[\Psi^*,\Psi]&=&{\lambda^2e^2\over2}
\sum_{m,n=1}^N\int_{\mb{x},\mb{y}}\Psi^*_{m,\mb{x}}\Psi_{m,\mb{x}}
{1\over4\pi|\mb{x}-\mb{y}|}\Psi^*_{n,\mb{y}}\Psi_{n,\mb{y}},\nonu
{\cal H}^\mr{i}[\Psi^*,\Psi,\rho]
&=&{1\over\beta}\rt\cdot{\delta\cC[\rho]\over\delta\rho}
-{1\over2\beta}\left({\delta\cC[\rho]\over\delta\rho}
-\rt\cdot{\lambda^2e^2\over\Delta}\right)\cdot\tG\cdot
\left({\delta\cC[\rho]\over\delta\rho}
-{\lambda^2e^2\over\Delta}\cdot\rt\right).
\eea
The variational form \eq{vareveq} has the following remarkable features. 
The generalized Hartree-Fock functional includes higher order radiative 
corrections which involve time-dependent, dynamical quantities. Furthermore,
it shows that the evolution, imposed at the ground state where the effective 
action is the best approximated, automatically involves an optimization
with respect to the choice of the quasi-particles, the wave functions
$\Psi_n$ in the electron field operator \eq{sqop}.

We shall consider two simple approximation schemes for the evolution equation
\eq{vareveq}: In a $c_2$ truncation we keep the $\ord{e^2}$ and $\ord{\rho^2}$
perturbative effective action and recover the usual Hartree-Fock equations
for the single particle wave functions $\Psi_n$ and energies $E_n$. 
Another $f_1c_2$ approximation scheme is where
one keeps the local potential $U(\rho)$ arbitrary but sets $\gamma=0$.
It is important to realize that the transformation
\be\label{usigmrt}
U(x,\rho)\to U(x,\rho)+\eta(x,\rho),~~~
\sigma(x,\rho)\to\sigma(x,\rho)+{2\over D_{x,x}}\eta(x,\rho)
\ee
leaves the effective action invariant. As pointed out above the photon
fluctuation determinant gives the exchange contribution and is
density independent. But the transformation \eq{usigmrt} can be used to
trade a possible, non-perturbative density-dependent photon self-energy 
term $\sigma$ into a local potential $U$. The result is a
generalization of the Kohn-Sham scheme \cite{ksh}, namely Eqs. \eq{vareveq} 
applied for the functional
\be
{\cal H}_\mr{KS}[\Psi^*,\Psi,E,\rho]=
{\cal H}^\mr{fr}[\Psi^*,\Psi,E]
+{\cal H}^\mr{C}_\mr{KS}[\Psi^*,\Psi,\rho]
+{\cal H}^\mr{i}_\mr{KS}[\Psi^*,\Psi,\rho],
\ee
where 
\bea
{\cal H}_\mr{KS}^\mr{C}[\Psi^*,\Psi,\rho]&=&-{n_s\lambda^2e^2}
\int_{\mb{x},\mb{y}}\sum_{n=1}^N\Psi^*_{n,\mb{x}}\Psi_{n,\mb{x}}
{1\over4\pi|\mb{x}-\mb{y}|}\rho_{\mb{y},0}\\
{\cal H}^\mr{i}_\mr{KS}[\Psi^*,\Psi,\rho]
&=&-{1\over2\beta}\rho\cdot{\lambda^2e^2\over\Delta}\cdot\tG\cdot
{\lambda^2e^2\over\Delta}\cdot\rho.
\eea
The ground state energy is determined by the differential equation
\bea\label{cev}
\partial_\lambda\int_xU(x,0)&=&-\lambda e^2 \tr\ \tG\ide
-\hf\rt\cdot{\lambda^2e^2\over\Delta}
\cdot\tG\cdot{2\lambda e^2\over\Delta}\cdot\tG\cdot
{\lambda^2e^2\over\Delta}\cdot\rt
\eea
together with the initial condition $U=0$ and contains the exchange term.

\section{Quenched averages}
The impurities represent a challenge in Condensed Matter Physics
since they appear to be static from the point of view of the
measurements and are distributed randomly. The usual way to 
take them into account is to average the connected Green functions, the
logarithm of the partition function over the impurity distributions \cite{edw}.
This average can be obtained either by analytical continuation
in the number of replicas \cite{replica}, or by the introduction
of fictious particles related to the real ones by super-transformations
\cite{susy} or by using the Keldysh contour in computing
loop-integrals \cite{keldysh}. The functional renormalization group
idea offers an alternative algorithm to compute quenched averages
which goes further than these methods in being fully non-perturbative
and in allowing annealed interactions \cite{cdf}. For the sake of 
simplicity we constrain the present discussion to the non-interacting case 
only.

Consider the generator functional for the Green functions of
the density and current, introduced in Eq. \eq{congrgf}, and
write its quenched average as
\be
\tilde W[\sigma]={\int\cd{v}e^{-{1\over2g}\int_\mb{x}v^2_\mb{x}}
W[\sigma_0+v,\sigma_1,\ldots]
\over\int\cd{v}e^{-{1\over2g}\int_\mb{x}v^2_\mb{x}}}.
\ee
The Legendre transform of $\tilde W[\sigma]$, the effective action $\Gamma[\rho]$, 
satisfies the evolution equation
\be
\partial_\lambda\Gamma[\rho]
=-{g\over2}\int_\mb{z}\left[{\delta^2\Gamma[\rho]\over
\delta\rho\delta\rho}\right]^{-1}_{(0,0,\mb{z}),(0,0,\mb{z})},
\ee
where the parameter $\lambda$ was introduced by making the rescaling
$g\to g\lambda$ and the notation of multiple-index $(\alpha,t,\mb{x})$ 
is used.

The gradient expansion ansatz for the functionals $\Gamma[\rho]$
seems natural when the kinetic transport coefficient 
are sought,
\bea\label{gradexp}
\Gamma[\rho]&=&\int_z\Biggl\{\hf\rho_{0,z}\left[
\partial_\omega\Gamma^{tt}i\partial_0
-\hf\partial^2_\omega\Gamma^{tt}\partial_0^2
-\partial_{q^2}\Gamma^{tt}\Delta\right]\rho_{0,z}\nonu
&&-{i\over2}\rho_{0,z}
(\Gamma^{ts}+\partial_\omega\Gamma^{ts}i\partial_0)\partial_k\rho_{k,z}
-{i\over2}\rho_{k,z}
(\Gamma^{st}+\partial_\omega\Gamma^{st}i\partial_0)\partial_k\rho_{0,z}\nonu
&&+\hf\rho_{j,z}\biggl[\delta_{j,k}\left(\Gamma^{ss}
+\partial_\omega\Gamma^{ss}i\partial_0
-\hf\partial^2_\omega\Gamma^{ss}\partial_0^2\right)\nonu
&&-\Gamma_T(\delta^{j,k}\Delta-\partial_j\partial_k)
-\Gamma_L\partial_j\partial_k\biggr]\rho_{k,z}
+{\vec{\rho}^2\over2}\Gamma^{ss}-\tilde U(\rho_0)\Biggr\}.
\eea
The $\Gamma$ coefficients are functions of the control parameter $\lambda$ 
and the density $\rho_0$ and have been computed in the leading order
of the perturbation expansion in $g$. Such an approximation is reliable 
for weak disorder, $\kappa=gm^2/2\pi\hbar^3p_F<1$, and one finds the form 
$P(\kappa)/p_F^n$ for the different $\Gamma$-functions where $P(\kappa)$ 
is simple regular function, $n>0$ and $\Gamma\to0$ when $\kappa\to\infty$.
Some combinations of the $\Gamma$-functions appear in the Kubo formula,
e.g. the electric conductivity,
$\sigma_{j,k}=\delta_{j,k}e^2\partial_\omega\Gamma^{ss}/(\Gamma^{ss})^2$,
the diffusion constant,
$D=\partial_{q^2}\Gamma^{tt}/(\Gamma^{tt})^2$, and the quantities
$\Gamma^{ss}$ and $\Gamma^{tt}=-\partial^2_{\rho_0}\tilde U$
give the density and current susceptibilities, respectively

When the vertex corrections are ignored then the evolution equation can 
be written as 
\bea
\partial_\lambda\Gamma&=&0~~~
\mr{for}~~~\Gamma=\partial_\omega\Gamma^{tt},\partial_\omega\Gamma^{ss},\nonu
\partial_\lambda\Gamma&=&g\partial^2_{\rho_0}\Gamma\int_\mb{p}{1\over2d^2_p}~~~
\mr{for}~~~\Gamma=\Gamma^{ss},\Gamma_T,\Gamma_L,\nonu
\partial_\lambda\gamma&=&g(\phi\partial_{\rho_0}+1)
\left(\partial_{\rho_0}^2\gamma\int_\mb{p}{1\over2d^2_p}\right)~~~
\mr{for}~~~\gamma=\partial^2_\omega\gamma^{tt},\partial_{q^2}\gamma^{tt},\nonu
\partial_\lambda\gamma&=&g(\phi\partial_{\rho_0}+2)\left(\partial^2_{\rho_0}\gamma
\int_\mb{p}{1\over8d^2_p}\right),~~~
\mr{for}~~~\gamma=\gamma^{ts},\partial_\omega\gamma^{ts},\nonu
\partial_\lambda\Gamma^{tt}&=&g\left(\partial^2_{\rho_0}\Gamma^{tt}
\int_\mb{p}{1\over2d^2_p}+\partial_{\rho_0}^2\partial_{q^2}\gamma^{tt}
\int_\mb{p}{\mb{p}^2\over2d^2_p}\right),
\eea
where
$\partial^2_\omega\gamma^{tt}=(\rho_0\partial_{\rho_0}+1)\partial^2_\omega\Gamma^{tt}$,
$\partial_{q^2}\gamma^{tt}=(\rho_0\partial_{\rho_0}+1)\partial_{q^2}\Gamma^{tt}$,
$\gamma^{ts}=(\rho_0\partial_{\rho_0}+2)\Gamma^{ts}$,
$\partial^2_\omega\gamma^{ts}=(\rho_0\partial_{\rho_0}+2)\partial^2_\omega\Gamma^{ts}$,
and
\be
d_p=\mb{p}^2(\rho_0\partial_{\rho_0}+1)\partial_{q^2}\Gamma^{tt}
+\Gamma^{tt}+{1\over4}\left[(\rho_0\partial_{\rho_0}+1)\Gamma^{ts}
-\Gamma^{st}\right]^2{\mb{p}^2\over\mb{p}^2\Gamma_L+\Gamma^{ss}}.
\ee
It is easy to see that the integration of these equations resums all 
correlation insertion of the particle-hole loop diagram in the given
truncation of the gradient expansion, in particular the maximally
crossed diagrams which are responsible for the cooperon pole needed for
non-vanishing conductivity \cite{cooperon}.

The terms $\ord{\partial_{\rho_0}^2}$ of the evolution equations correspond
to a formal diffusion process on the plane $(\rho_0,g)$ considered as 
space-time. One can
verify that the remaining terms which contain $\rho_0\partial_{\rho_0}^3$
generate similar spread, as well. The diffusion constant in the $\rho_0$-space,
the integral on the right hand sides, is small at the initial condition for 
strong disorder and becomes large for weak disorder. 
Therefore the $\Gamma$-functions will approach zero for weak disorder 
due to the strong diffusion. This is supposed to generate
weak localization. 

The localization transition should occur when some or all $\ord{\partial^2}$
terms of the effective action are vanishing. In fact, according to the
Kubo formulae the mobility is proportional to certain $\ord{\partial^2}$
terms in the effective action. Another reasoning is to recall the reduction
formulae for a single particle which gives the scattering amplitudes in terms 
of the residuum of the connected propagator on the mass shell. It remains to
be seen by detailed numerical studies of the evolution equation in 3 dimensions
whether the $\rho_0$-dependence of the integrals on the right hand sides 
provides such a self-acceleration of the diffusion process in the 
$\rho_0$-space which cancels some $\Gamma$-functions at finite value of $g$.

We note finally that there is a formal analogy between localization and
the phenomenon of the quark confinement. According to the haaron-model
of the QCD vacuum the quark propagator is $\ord{p^{-4}}$ and the quark 
confinement appears as a localization in the space-time \cite{haaron}.

\section{Summary}
Two applications of the internal space renormalization group were given
demonstrating the possibility of a new, systematical non-perturbative
method to tackle the bound state problem at least when the bound state
formation is triggered by an external potential. 

This method is in its infancy and the outline of its formal structure 
is sketched only.
But we believe that it can provide an accuracy and flexibility superior 
to other procedures when formal computer algebra and numerical integration
are combined in deriving and solving the evolution equation in a sufficiently
rich functional space.

\end{document}